\documentclass[sigconf]{acmart}
\usepackage{CJKutf8}
\usepackage{booktabs}
\usepackage{multirow}
\usepackage{graphicx}
\usepackage{multirow}
\usepackage{makecell}
\usepackage{caption}
\usepackage{graphicx} 
\usepackage{float} 
\usepackage{subfigure} 
\usepackage{caption}
\usepackage{booktabs}
\usepackage{multirow}
\AtBeginDocument{%
  \providecommand\BibTeX{{%
    \normalfont B\kern-0.5em{\scshape i\kern-0.25em b}\kern-0.8em\TeX}}}

\copyrightyear{2022} 
\acmYear{2022} 
\setcopyright{rightsretained}
\acmConference[DDAM '22]{Proceedings of the 1st International Workshop on Deepfake Detection for Audio Multimedia}{October 14, 2022}{Lisboa, Portugal}
\acmBooktitle{Proceedings of the 1st Int'l Workshop on Deepfake Detection for Audio Multimedia (DDAM '22), Oct. 14, 2022, Lisboa, Portugal}\acmDOI{10.1145/3552466.3556525}
\acmISBN{978-1-4503-9496-3/22/10}

\begin{document}
\title{An Initial Investigation for Detecting Vocoder Fingerprints of Fake Audio}


%\author{Anonymous authors}
% \affiliation{
%   \institution{Paper under double-blind review}
% }
%\renewcommand{\shortauthors}{Anonymous Author, et al.}

\author{Xinrui Yan}
%\email{yanxinrui2021@ia.ac.cn}
\affiliation{%
  \institution{Institute of Automation, Chinese Academy of Sciences}
  \institution{University of Chinese Academy of Sciences}
  \city{Beijing}
  \country{China}
}

\author{Jiangyan Yi}
%\email{yanxinrui2021@ia.ac.cn}
\affiliation{%
%  \institution{School of Artificial Intelligence, University of Chinese Academy of Sciences, China}
  \institution{Institute of Automation, Chinese Academy of Sciences}
  \institution{University of Chinese Academy of Sciences}
  \city{Beijing}
  \country{China}
 %\country{China}
}

\author{Jianhua Tao}
%\email{yanxinrui2021@ia.ac.cn}
\affiliation{%
  \institution{Institute of Automation, Chinese Academy of Sciences}
  \institution{University of Chinese Academy of Sciences}
  %\institution{CAS Center for Excellence in Brain Science and Intelligence Technology}
  \city{Beijing}
  \country{China}
 %\country{China}
}

\author{Chenglong Wang}
%\email{yanxinrui2021@ia.ac.cn}
\affiliation{%
  \institution{Institute of Automation, Chinese Academy of Sciences}
  \institution{University of Science and Technology of China}
%  \institution{Center for Excellence in Brain Science and Intelligence Technology, China}
%\streetaddress{P.O. Box 1212}
 \city{Beijing}
 \country{China}
}

\author{Haoxin Ma}
%\email{yanxinrui2021@ia.ac.cn}
\affiliation{%
  \institution{Institute of Automation, Chinese Academy of Sciences}
  \institution{University of Chinese Academy of Sciences}
%\streetaddress{P.O. Box 1212}
 \city{Beijing}
 \country{China}
}

\author{Tao Wang}
%\email{yanxinrui2021@ia.ac.cn}
\affiliation{%
  \institution{Institute of Automation, Chinese Academy of Sciences}
  \institution{University of Chinese Academy of Sciences}
%\streetaddress{P.O. Box 1212}
 \city{Beijing}
 \country{China}
}

\author{Shiming Wang}
\affiliation{%
  \institution{Institute of Automation, Chinese Academy of Sciences}
  \institution{University of Science and Technology of China}
%\streetaddress{P.O. Box 1212}
 \city{Beijing}
 \country{China}
}

\author{Ruibo Fu}
\affiliation{%
  \institution{Institute of Automation, Chinese Academy of Sciences}
  %\institution{University of Chinese Academy of Sciences}
%\streetaddress{P.O. Box 1212}
 \city{Beijing}
 \country{China}
}

\renewcommand{\shortauthors}{Xinrui Yan and Jiangyan Yi, et al.}

\begin{abstract}
Many effective attempts have been made for fake audio detection. However, they can only provide detection results but no countermeasures to curb this harm. For many related practical applications, what model or algorithm generated the fake audio also is needed. Therefore, We propose a new problem for detecting vocoder fingerprints of fake audio. Experiments are conducted on the datasets synthesized by eight state-of-the-art vocoders. We have preliminarily explored the features and model architectures. The t-SNE visualization shows that different vocoders generate distinct vocoder fingerprints.
\end{abstract}

\begin{CCSXML}
<ccs2012>
   <concept>
       <concept_id>10010405.10010462.10010466</concept_id>
       <concept_desc>Applied computing~Network forensics</concept_desc>
       <concept_significance>500</concept_significance>
       </concept>
 </ccs2012>
\end{CCSXML}

\ccsdesc[500]{Applied computing~Network forensics}

\keywords{fake audio; vocoder fingerprints detection; audio forensics}


\maketitle

\section{Introduction}
With advances in deep learning and fake technologies, synthetic speech is getting closer and closer to a natural voice. Some of the most state-of-the-art text-to-speech (TTS) technologies \cite{ze2013statistical} \cite{qian2014training} \cite{zen2015unidirectional} \cite{wang2016first} \cite{li2018emphasis} \cite{oord2016wavenet} achieve such a high degree of naturalness that even humans have difficulty distinguishing between real speech and fake speech.

\begin{figure*}[ht]
  \centering
  \includegraphics[width=\textwidth, height=8cm]{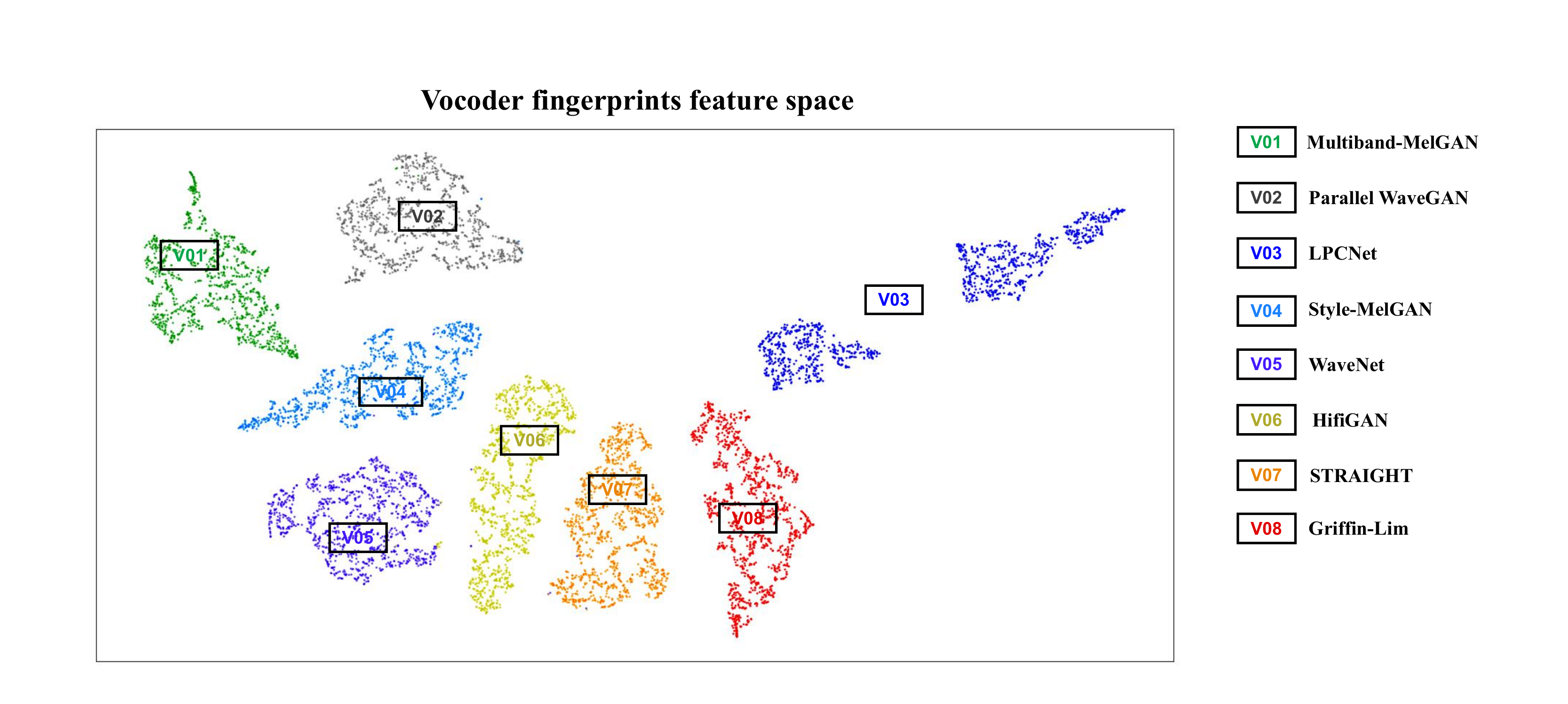}
  \caption{A t-SNE \cite{van2008visualizing} visual of vocoder fingerprints features of fake audio. In the vocoder fingerprints feature space, different colors represent different vocoders: V01 represents Multiband-MelGAN, V02 represents Parallel WaveGAN, V03 represents LPCNet, V04 represents Style-MelGAN, V05 represents WaveNet, V06 represents HifiGAN, V07 represents STRAIGHT, V08 represents Griffin-Lim. The visualization result shows that different text-to-speech (TTS) systems can result in distinct vocoder fingerprints features.}
  \label{fig:teaser}
\end{figure*}

Unfortunately, they can also be used as a powerful tool to spread misinformation. For example, spreading false political narratives, stealing audio rights from vendors, synthesizing malicious audio to defame somebody, etc. Given the significant increase in the generation and ease of access to fake multimedia content, it is becoming increasingly difficult to know the truth. This has posed a significant threat to social stability and security. Therefore, fake audio detection has great significance. 

Fake audio detection has made an increasing number of attempts \cite{wang2021investigating} \cite{yi2021half} \cite{tak2021end} \cite{ma2021continual} \cite{yi2022add}, which are broadly divided into two categories: (i) feature engineering at the front end and (ii) classifier design at the back end. Previous literature reported that several features are effective for fake detection. The advantage of the variable resolution of Constant-Q Cepstral Coefficients (CQCCs) \cite{todisco2016new} has been widely recognized. The common Mel-Frequency Cepstral Coefficients (MFCCs) \cite{davis1980comparison} are also used in replay attack detection. \cite{lawson2011survey} found that the Linear Frequency Cepstral Coefficients (LFCCs) improved  accuracy over MFCCs in many conditions.

As for the back-end classifier, effective classifier designs are considered to be more robust. As a baseline for the ASVspoof challenge \cite{wu2015asvspoof} \cite{kinnunen2017asvspoof} \cite{todisco2019asvspoof} \cite{yamagishi2021asvspoof}, the Lightweight Convolutional Neural Network (LCNN) \cite{bagherinezhad2017lcnn} is used widely in the field of fake audio detection. LCNN can not only separate noise and information signals but also can select features through competitive learning. To further alleviates the problem of gradient disappearance, He et al. \cite{he2016deep} introduced residual network (ResNet). Hu et al. \cite{hu2018squeeze} further proposed SE-ResNet. The "Squeeze-and-Excitation" (SE) block is introduced to explicitly model channel interdependencies and adaptively recalibrate the characteristic response at the channel level. In recent years, X-vectors \cite{snyder2018x} have consistently provided state-of-the-art results on the task of speaker verification. The X-vector architecture is a time-delayed neural network (TDNN) \cite{waibel1989phoneme}, which applies a statistical pool to project a variable-length corpus into a fixed-length embedding of speaker features. 

All of the above work has made many contributions to fake audio detection. However, All these works focus on distinguishing between the two categories of real audio and fake audio. For fake audio, there is no further explainability work and effective countermeasures to curb this security issue.

However, in many application scenarios (for more details refer to Section 2), not only do they care about the authenticity of the audio itself, but also need to know what model or algorithm generated it. That is, if the audio is detected as fake audio, we also want to know which model generated it. For example, fake audio is from HifiGAN vocoder, etc. Therefore, it is crucial to find effective ways to detect which model generated the fake audio.

With the development of deep learning, TTS has significantly improved the quality of fake audio in recent years. A modern TTS system consists of three basic components: a text analysis module, an acoustic model, and a vocoder. The text analysis module converts the text sequence into linguistic features. The acoustic model generates acoustic features from the linguistic features, and then the vocoders synthesize the waveform from the acoustic features. The vocoder is one of the important components, and a large amount of research work comes out focusing on different vocoders of TTS. In general, the development of vocoders can be categorized into two stages: pure signal-processing-based vocoders, and the neural network-based vocoders. Some popular vocoders in SPSS include STRAIGHT \cite{kawahara2006straight} and WORLD \cite{morise2016world}. The neural vocoders are into different categories: autoregressive vocoders such as LPCNet \cite{valin2019lpcnet}, WaveNet \cite{oord2018parallel} and WaveRNN \cite{kalchbrenner2018efficient}, Flow-based vocoders, GAN-based vocoders such as Parallel WaveGAN \cite{yamamoto2020parallel}, HifiGAN \cite{kong2020hifi}, Multiband-MelGAN\cite{yang2021multi}, Style-MelGAN\cite{mustafa2021stylemelgan}, etc. In addition, there are traditional vocoders such as Griffin-Lim \cite{griffin1984signal}, etc. Besides, among the methods used for state-of-the-art speech synthesis, the vocoders is the last step to creating fake waveform, and they leave traits in the synthesized waveform and are potential to be detected. So fake audio synthesized by different vocoders poses serious risks. This makes it necessary to explore methods for different vocoders to detect the generation source of fake audio. For these reasons, the focus of this work is on detecting vocoder fingerprints of fake audio.

To solve this problem, we proposed a new problem: detecting vocoder fingerprints of fake audio. This paper initially focuses on an investigation for detecting vocoder fingerprints of fake audio. In this paper, vocoder fingerprints are different characteristics of audio from different vocoders of TTS systems. We construct a vocoder fingerprints detection datasets that include fake audio synthesized by eight different vocoders. With the datasets, we performed a contrast analysis of vocoder fingerprints from the eight vocoders including STRAIGHT, LPCNet, WaveNet, PWG, HifiGAN, Multiband-MelGAN, Style-MelGAN, Griffin-Lim. The results show that different vocoder fingerprints show excellent distinguishability. And LFCC features perform best among all features and the ResNet model achieved the highest detection rate. We explicitly visualize the vocoder fingerprints in the fake audio to better explain the validity, as shown in Figure 1.

\begin{figure*}[ht]
  \centering
  \includegraphics[width=\textwidth]{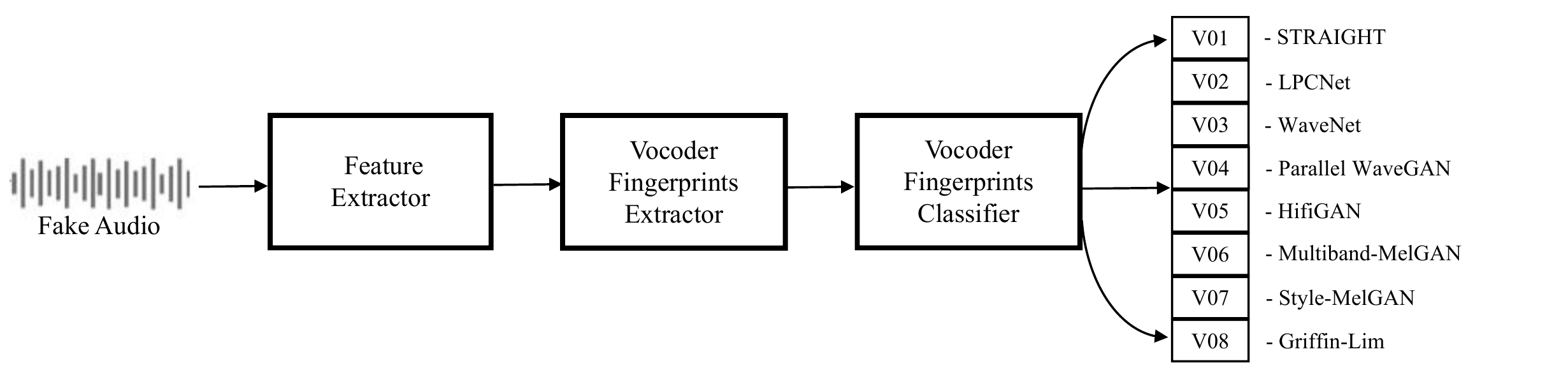}
  \caption{The overall framework of vocoder fingerprints detection.}
 
\end{figure*}

The main contributions of this paper are as follows. 
\begin{itemize}
\item To the best of our knowledge, we propose the concept of vocoder fingerprints for the first time, and list many application scenarios of detecting vocoder fingerprints.
\end{itemize}
\begin{itemize}
\item We research how to detect vocoder fingerprints and analyze the differences in fingerprints of different vocoders. Based on the construction of vocoder fingerprints detection datasets, we provided strong benchmarks for other researchers so that researchers can propose better methods.
\end{itemize}
\begin{itemize}
\item For the task of vocoder fingerprints detection, we discuss the limitations of current detection methods and possible research directions in the future.
\end{itemize}
The rest of the paper is organized as follows. Section 2 introduces some related applications. Section 3 explains our method for detecting vocoder fingerprints. Section 4 gives the evaluation metric. Besides, experiments, discussion and future work are reported in Section 5 and Section 6 respectively. Finally the conclusions are given in Section 7.

\section{Related applications}

Our work in this field mainly, but not exclusively, includes these important applications. The details are listed as follows.

\textbf{Audio forensics and multimedia forensics.}
Audio is one medium for multimedia. The work of this paper contributes to audio forensics, but it is also helpful to multimedia forensics. It is difficult to effectively evaluate and analyze video only using one medium for multimedia forensics. Multimedia forensics or video forensics \cite{milani2012overview} \cite{stamm2012temporal} benefits from the combination and collaboration of audio, video, text, etc.

\textbf{Intellectual property protection.}
As virtual property, virtual or synthetic media should be protected \cite{yu2019attributing}. Copyright plagiarism of fake audio may endanger the intellectual property of the company's system owners. Therefore, fake audio forensics is very desirable for the protection of intellectual property. Manufacturers can also detect whether their synthetic audio is stolen.

\textbf{Judicial evidence appraisal.}
In every department of judicial proceedings (such as the Public Security Bureau or the Court), it is normal to use audio recordings as evidence. Audio recognized as evidence must be authentic and its integrity must be verified. If the fake audio technology is used to forge evidence in the judicial field, it will cause great harm. If it is detected that the evidence is fake, further evidence collection based on the detection can provide an explanation, to achieve the purpose of promoting judicial justice. This is expected to be a challenging task.

\textbf{Explainability of fake detection results.}
In recent years, fake poses a great threat to social stability and security. Fake audio detection is an emerging topic and an increasing number of attempts have been made. However, it is also necessary to develop dedicated countermeasures to solve the security problem. One of the methods is to explain the detection results. It is crucial to decide which generation method synthesized the fake audio, which will help us find the source of the fake audio.

\section{Vocoder fingerprints detection}
The overall architecture of the proposed vocoder fingerprints detection is shown in Figure 2. It is composed of three major parts: acoustic feature extractor, vocoder fingerprints extractor, and vocoder fingerprints classifier. First, LFCC features are extracted from the feature extractor, then LFCCs are fed into the vocoder fingerprints extractor of ResNet to get all the vocoder fingerprints. Finally, the vocoder fingerprints are fed into the fully-connected classifier for classification to realize the forensics of eight types of vocoders.

\renewcommand\arraystretch{1.2}
\begin{table*}[ht]
\footnotesize
\caption{Number of utterances and speakers for training, 
development, and test sets}
\small
\resizebox{\linewidth}{!}{
\begin{tabular}{@{}c|c|ccccccccc@{}}
\toprule
                             &              & STRAIGHT & LPCNet & WaveNet &  Parallel WaveGAN  & HifiGAN & Multiband-MelGAN & Style-MelGAN & Griffin-Lim & Total \\ \midrule
\multirow{2}{*}{Train}       & \#Utterances & 3200     & 3200   & 3200    & 3200 & 3200    & 3200             & 3200         & 3200        & 25600 \\
                             & \#Speakers   & 40       & 40     & 40      & 33   & 40      & 33               & 33           & 40          & 299   \\ \midrule
\multirow{2}{*}{Development} & \#Utterances & 1200     & 1200   & 1200    & 1200 & 1200    & 1200             & 1200         & 1200        & 9600  \\
                             & \#Speakers   & 20       & 13     & 20      & 13   & 9       & 13               & 13           & 13          & 114   \\ \midrule
\multirow{2}{*}{Test}        & \#Utterances & 3500     & 3500   & 3500    & 3500 & 3500    & 3500             & 3500         & 3500        & 28000 \\
                             & \#Speakers   & 50       & 30     & 50      & 24   & 47      & 24               & 24           & 30          & 279   \\ \bottomrule
\end{tabular}}
\end{table*}
\subsection{Feature extraction}

The extracted Linear Frequency Cepstrum Coefficients (LFCCs) are based on the cepstrum features of triangular filter banks. LFCC features select linear filter banks, so that the extracted features have the same resolution for both low and high frequency regions. The LFCCs extraction process includes signal pre-emphasis, framing, windowing, Fast Fourier Transform (FFT), power spectrum calculation, passing through the linear filter banks, logarithmic operation, Discrete Cosine Transform (DCT), and differential calculation, etc. Finally, static features with their first-order difference and second-order difference coefficients to obtain 60-dimensional LFCC features. After acoustic feature extraction, the extracted features are fed into the vocoder fingerprints extractor.

\subsection{Model architecture}

In this paper, we use ResNet as our model for vocoder fingerprints detection. It is well known that the residual module proposed by ResNet largely alleviates the problem of gradient disappearance. ResNet introduces a shortcut connection that allows the gradient to flow through a large number of layers. In our experiments, we use a residual network with eight basic blocks. The structure of each basic block has two 3x3 convolutional layers, both with 16 channels. We consider a basic block to be defined as:

\begin{eqnarray}
y & = & F\left ( x, \left \{W_{i}  \right \} \right ) +  x
\end{eqnarray}

Here $x$ and $y$ represent the input and output vectors. $F\left (x,  \left \{ W i\right \} \right )$ represents the residual mapping to be learned. When changing the input/output channels, a linear projection $W_s$ can be used to match the dimensions:

\begin{eqnarray}
y & = & F\left ( x,\left \{ W_{i}  \right \}  \right ) +  W_{s}x 
\end{eqnarray}

If $H(x)$ represents an underlying mapping. According to equation 2, in the case where the input and output have the same dimension, multiple nonlinear layers can asymptotically approximate the residual functions, i.e., $H(x) - x$. For $F(x) = H(x) - x$, let these layers approximate a residual function $F(x)$. Thus, the original mapping to be learned becomes $F(x)+x$.

The ResNet-18 we use consists of layers in the following order: 7×7 convolutional layers, maxpool layer, eight basic blocks, global average pooling layer.

Using these overall network architecture above as a vocoder fingerprints extractor, the output of the last layer mentioned above are our desired vocoder fingerprints. The extracted vocoder fingerprints finally input the fully connected classification layer, and the final output corresponds to eight vocoders: STRAIGHT, LPCNet, WaveNet, Parallel WaveGAN, HifiGAN, Multiband-MelGAN, Style-MelGAN, Griffin-Lim.

\section{EVALUATION METRIC}

The performance of our forensic countermeasures is evaluated via $Precision$, $Recall$ and $F_1-score$ \cite{chicco2020advantages}. In detection problems, systems are designed to decide whether a given event or feature is present or absent in a given space. Quantifying a system performance is normally done by combining True/False Positives/Negatives to measure the $Precision$ and $Recall$. $Precision$ and $Recall$ are defined by:

\begin{eqnarray}
Precision & = & \frac{TP}{TP+FP}  
\end{eqnarray}

\begin{eqnarray}
Recall & = & \frac{TP }{TP+FN } 
\end{eqnarray}

where $TP$ , $FP$ and $FN$ denote the true positive, false positive, false negative, respectively.
Finally, the $F_1-score$ can be calculated by:

\begin{eqnarray}
F_{1}-score & = & 2\times \frac{Precision \times Recall }{Precision +Recall } 
\end{eqnarray}

\section{EXPERIMENTS}

\subsection{Datasets}

The vocoder fingerprints detection datasets consists of 25600 training utterances, 9600 development utterances and 28000 test utterances. The datasets are built from the freely available TTS systems and generated using AISHELL3 chinese corpus. All fake audios were synthesized separately by eight vocoders, including STRAIGHT \footnote{$https://github.com/HidekiKawahara/legacy\_STRAIGHT.git$}, LPCNet \footnote{https://github.com/xiph/LPCNet.git}, WaveNet \footnote{https://github.com/r9y9/wavenet\_vocoder.git}, Parallel WaveGAN \footnote{https://github.com/kan-bayashi/ParallelWaveGAN.git
\label{url:pwg}}, HifiGAN \textsuperscript{\ref{url:pwg}}, Multiband-MelGAN \textsuperscript{\ref{url:pwg}}, Style-MelGAN \textsuperscript{\ref{url:pwg}}, Griffin-Lim \footnote{https://librosa.org/}. The acoustic model is fixed. The general composition of the datasets is shown in Table 1, showing the number of utterances and speakers of each vocoder. The training set consists of 299 speakers, the development set consists of 114 speakers, and the test set consists of 279 speakers. There was no overlap of speakers between these three datasets.

\renewcommand\arraystretch{1.2}
\begin{table*}[ht]
\caption{With different feature, the $F_1$-score (\%) values of different models trained on different vocoders.}
\resizebox{\linewidth}{!}{
\begin{tabular}{@{}c|c|ccccccccc@{}}
\toprule
                             &              & STRAIGHT & LPCNet & WaveNet & Parallel WaveGAN  & HifiGAN & Multiband-MelGAN & Style-MelGAN & Griffin-Lim & Total \\ \midrule
\multirow{4}{*}{CQCC}       & X-vector &60.72 	&88.54 	&80.19 	&68.81 	&66.54 	&58.44 	&92.40 	&99.91 	&76.95  \\
                             & LCNN  &99.90  &100.00	&99.97 	&98.69 	&85.20 	&75.94 	&99.21 	&100.00 	&94.86 \\
                             & SE-ResNet   &99.97 	&100.00  &99.99 	&97.08 	&91.36 	&84.49 	&99.01  &100.00 	&96.49 
                              \\
                             & ResNet   &99.81 	&100.00 	&99.97 	&98.54 	&90.05 	&85.25 	&99.80 	&100.00 	&\textbf{96.68} 
                             \\\midrule
\multirow{4}{*}{MFCC}       & X-vector &92.78 	&95.12 	&93.81 	&82.04 	&94.94 	&74.86 	&91.94 	&97.08 	&90.32   \\
                             & LCNN   &99.36 	&99.89 	&98.91 	&85.31 	&99.77 	&71.17 	&95.16 	&99.63 	&93.65 \\
                             & SE-ResNet    &99.44 	&99.96 	&96.99 	&95.28 	&99.64 	&70.58 	&86.44 	&99.86 	&93.52 
                              \\
                             & ResNet   & 98.80 	&99.77 	&98.66 	&96.65 	&99.79 	&94.96 	&98.59 	&99.56 		&\textbf{98.35}
                             \\\midrule
\multirow{4}{*}{LFCC}       & X-vector & 99.91 	&100.00 	&99.67 	&81.89 	&99.76 	&70.72 	&98.31 	&99.54 	&93.72 \\
                             & LCNN   &100.00 	&100.00 	&99.96 	&92.02 	&98.16 	&88.09 	&99.86 	&100.00 	&97.26\\
                             & SE-ResNet   &  99.96 	&100.00 	&99.99 	&99.36 	&99.16 	&96.13 	&97.81 	&99.97 	&99.05
                              \\
                             & ResNet   &\textbf{100.00} 	&\textbf{100.00} 	&\textbf{99.99} 	&\textbf{100.00} 	&\textbf{99.99}	&\textbf{99.99} 	&\textbf{99.99} 	&\textbf{100.00} 	&\textbf{99.99}
                             \\\bottomrule

\end{tabular}
}
\end{table*}

% \usepackage{booktabs}
\begin{table*}[ht]
\centering
\normalsize
\caption{Comparison of the average $F_1$-score (\%) values of eight vocoders under three different feature.}
\small
\begin{tabular}{@{}c|ccccccccc@{}}
\toprule
        & STRAIGHT & LPCNet & WaveNet & Parallel WaveGAN & HifiGAN & Multiband-MelGAN & Style-MelGAN & Griffin-Lim & Total \\ \midrule
CQCC    & 90.10    & 97.14  & 95.03   & 90.78            & 83.29   & 76.03            & 97.61        & 99.98       & 91.25 \\
MFCC    & 97.60    & 98.69  & 97.09   & 89.82            & 98.54   & 77.89            & 93.03        & 99.03       & 93.96 \\
LFCC    & 99.97    & 100.00 & 99.90   & 93.32            & 99.27   & 88.73            & 98.99        & 99.88       & 97.51 \\
Average & 95.89    & 98.61  & 97.34   & 91.31            & 93.70   & 80.89            & 96.54        & 99.63       & 94.24 \\ \bottomrule
\end{tabular}
\end{table*}

\subsection{Experimental setup}

\textbf{Front-end feature extraction:}
For feature engineering exploration, comparative analysis of vocoder fingerprints detection was performed using CQCCs, MFCCs, and LFCCs, respectively. We obtained 60-dimensional CQCCs as low-level input. To extract MFCCs, we applied a Hamming analysis window, the size of which is 25ms with a 10ms shift, and we employed a mel-filter bank with 26 channels. In addition to that, We extract the 60-dimensional LFCCs. For LFCCs, the length of the Hamming analysis window is set to 25ms. The number of FFT bins is set to 512 and frame shift is set to 10ms. The number of filters is set to 20.

\textbf{Back-end model architecture:}
As for the back-end model architecture, We experimented with
X-vector, LCNN, SENet,and ResNet.

\begin{itemize}
\item \textbf{X-vector}: We adopted the X-vector speaker recognition system builds on TDNN embedding architecture. This repo contains the implementation of  \cite{snyder2018spoken} in Pytorch. The first five layers called frame layers operate at the sequential frame level.  At the statistical pooling layer, frame 5 outputs from all frames are aggregated by computing the mean and standard deviation. The subsequent frames operate on this 1024- dimensional vector which represents the entire segment and are named segment layers. Then we extract embeddings at segment6, the output dimension at this layer is set to 512-dimensional. The extracted embeddings are the x-vectors. The final output layer corresponds to the eight types of vocoders.
\end{itemize}

\begin{itemize}
\item \textbf{LCNN}: 
The LCNN model refers to \cite{lavrentyeva2019stc}, but the 28th layer adopts Adaptive MaxPool2d. The specific characteristic of Light CNN architecture is the usage of the Max-Feature-Map activation (MFM) which is based on Max-Out activation function.
\end{itemize}

\begin{itemize}
\item \textbf{SE-ResNet}: In SE-ResNet, the “Squeeze-and-Excitation” block adaptively recalibrates per channel feature responses by explicitly modelling interdependencies between channels. Each SE-ResNet module  is comprised of two convolutional layers with 3 × 3 kernels and one SE block. SE block is used to improve the representational capacity of the network by enabling it to perform dynamic channel-wise feature recalibration. The features input through the Adaptive Average-pooling layer. Then two fully connected layers are used to downsample and upsample. The final output of the SE block is the channel-wise multiplication.
\end{itemize}

\begin{itemize}
\item \textbf{ResNet}: The deep residual learning framework has fewer filters and lower complexity. We chose 18-layers ResNet that converges faster. For each building block, we use a stack of two convolutional layers. The convolutional layers are all with kernels of 3 × 3. The subsampling is performed by convolutions with a stride of 2. The network ends with a global average pooling, a fully-connected layer. Finally, the network outputs the classification results of the eight vocoders.
\end{itemize}

\textbf{Training strategy:} We implement all models with PyTorch. Parameters are initialized randomly. We use Adam optimizer with the $\beta_1$ parameter set to 0.9 and the $\beta_2$ parameter set to 0.98 to update the weights in the model. The initial learning rate is set to 0.001 with linear learning rate decay. These models are trained with a batch size of 256. The num epochs is set up 100.

\subsection{Performance of vocoder fingerprints detection with different features}
As shown in Table 2, the LFCC features are relatively more suitable for the task of vocoder fingerprints detection.
The test results show that the LFCC features get the best performance on all four models. The total $F_1$-score values are 93.72\%, 97.26\%, 99.05\%, 99.99\% respectively, and the average $F_1$-score value is 97.51\%.
As shown in Table 3, from the vocoder perspective, the LFCC features achieve the highest vocoder fingerprints detection rate on almost all the eight vocoders. The average $F_1$-score values are 99.97\%, 100.00\%, 99.90\%, 93.32\%, 99.27\% , 88.73\%, 98.99\%, and 99.88\%, respectively. Except for the Griffin-Lim, the $F_1$-score values on the other seven vocoders were higher than those on MFCC and CQCC.

The literature shows that LFCC has better resolution in the high frequency region as a more comprehensive linear feature. As shown in Figure 3, the audio synthesized through the eight vocoders differed more in the high frequencies. Therefore, this reason has the potential to make LFCC a relatively more suitable feature for detecting vocoder fingerprints.

Besides that, MFCC has a better detection result with an average $F_1$-score value of 93.96\%, higher than the average $F_1$-score value of 91.25\% for CQCC.

\begin{figure*}[ht]
  \centering
  \includegraphics[width=0.95\textwidth]{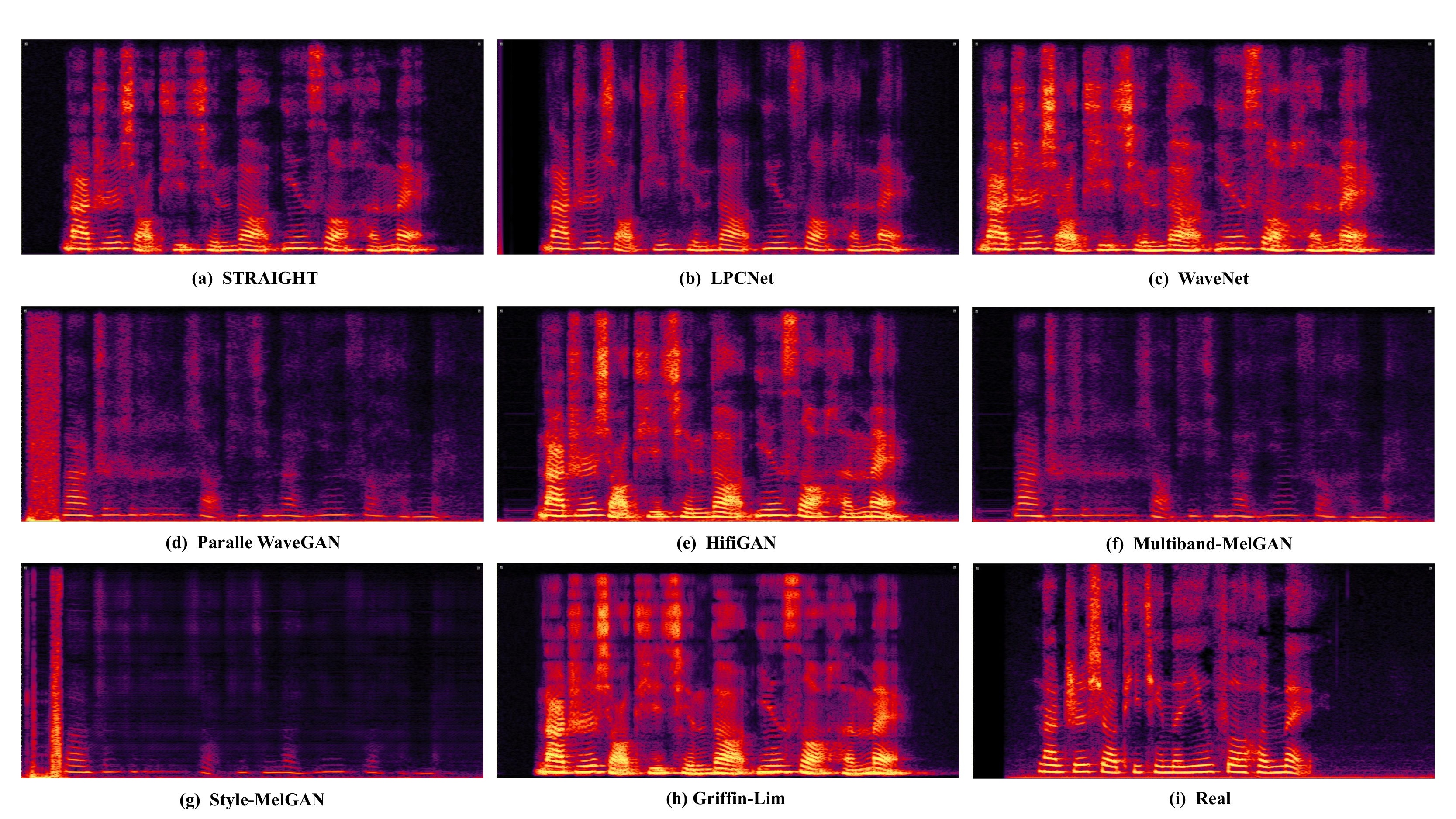}
  \caption{A spectrogram comparison of the fake audio through eights vocoders, the content of each audio is
  \begin{CJK*}{UTF8}{gbsn}
‘ 那只小提琴的音色很美. ' (‘ That violin has a beautiful tone. ')
  \end{CJK*}
 The abscissa of the spectrogram is time and the ordinate is frequency. The coordinate point value is speech data energy. The magnitude of the speech data energy value is indicated by color.}
  \Description{A woman and a girl in white dresses sit in an open car.}
\end{figure*}

\subsection{Performance of vocoder fingerprints detection with different model architectures}
As shown in Table 2, the ResNet model can better identify the vocoder fingerprints in all cases of trained with three features, achieving the total $F_1$-score values of 96.68\% and 98.35\%, 99.99\%, respectively. And the LFCC-ResNet detect system achieves the highest total $F_1$-score value of 99.99\% among all baseline results.
From the vocoder point of view, the ResNet model achieves the highest vocoder fingerprints detection rate on all eight vocoders. The $F_1$-score are: 100.00\% ,100.00\% ,99.99\% ,100.00\% ,99.99\%, 99.99\% ,99.99\% ,100.00\%, respectively. 

The relatively poor performance of X-vector may be that the goal of X-vector is focused on classifying a large number of speakers. However, this does not fit well with the goal of the vocoder fingerprints detection task. Whereas LCNN separates noisy signals from informative signals by competitive learning, however, our datasets audios are in a clean environment. The SE-ResNet network is to focus on the relationship between channels, hoping that the model can automatically learn more features of different channels. This advantage may not work well for our vocoder fingerprints detection.

\subsection{Comparison of the performance for detecting vocoder fingerprints of eight vocoders}

As shown in Table 3, we have some findings for the detection of eight types of vocoder fingerprints under three feature input environments. The Multiband-MelGAN's vocoder fingerprints are relatively the most difficult to detect. Griffin-Lim and LPCNet's vocoder fingerprints are relatively easier to be detected. We observed the comparative speech spectrograms through the eight vocoders with the same utterance (as shown in Figure 3), and found that Griffin-Lim's speech spectrogram has more missing information in the high-frequency region, a characteristic that is not found in other vocoders. In addition, LPCNet's initial phase retains audio information for a shorter period of time, while the initial phase of the other vocoders is almost all silent regions.

There is also an exception that HiFiGAN exhibits an unusual vocoder fingerprints detection effect when utilizing CQCC as the front-end input. Under both the MFCC and LFCC feature input environments, HiFiGAN's vocoder fingerprints are higher than Parallel WaveGAN and Multiband-MelGAN. In contrast, the vocoder fingerprints of HiFiGAN become difficult to detect under the features environment of CQCC. The specific reason for this deserves further investigation.

\section{DISCUSSION}
In this section, we discuss some of its limitations. The limitations suggest some areas for further research.

\begin{itemize}
\item \textbf{More diverse vocoder detection datasets:} 
The vocoder fingerprints dataset in this paper contains the current common eight vocoders, but this number is still limited for the vocoder forensics task. In the future, we will expand more vocoder types and waveform splicing types, etc. In addition, our datasets corpus is all from Chinese AISHELL3, and future work will further add English datasets and make the more diverse vocoder fingerprints datasets publicly available.
\end{itemize}

\begin{itemize}
\item \textbf{Noise robustness:}
The work here entirely focuses on clean fake audios without significant noise or channel effects. In order to make the proposed countermeasures suitable for practical applications, channel and noise issues must of course be taken into account. In the future, we will further investigate this under different signal to noise ratios and different scene noises to being able to better apply this task to real scenarios.
\end{itemize}

\begin{itemize}
\item \textbf{More innovative methods:} 
To the best of our knowledge, this is the first study to detect the vocoder fingerprints of fake audio. Our goal is to investigate whether the vocoder fingerprints can be distinguished and how to detect the vocoder fingerprints of fake audio. 
We have preliminarily explored the features and model architectures. Our method chose the methods used in fake detection tasks, which obtain promising results. We hope to provide strong benchmarks for other researchers so that researchers can propose better methods. In the future, we will propose more novel model architecture on our findings.
\end{itemize}

\begin{itemize}
\item \textbf{Truly generalized countermeasures:}
The proposed countermeasure does not apply well to unknown vocoders. In the future, we will conduct further research on the forensics of unknown types of vocoders to better achieve the generalizability of vocoder forensics.
\end{itemize}

\section{CONCLUSIONS}
In all existing literature that we are aware of in fake audio detection, the lack of interpretation of detection results and countermeasures to address this security issue is a  barrier to progress. We aim to investigate whether the vocoder fingerprints can be distinguished and how to detect the vocoder fingerprints of fake audios. The results show that LFCC features are relatively more suitable for vocoder fingerprints detection. Besides, the ResNet achieves the best detection results among other models. We hope to provide strong benchmarks for other researchers to facilitate more innovative methods. Future work includes considering improving the limitations of the discussion. 

\bibliographystyle{unsrt}
\bibliography{acmart}

\clearpage

\end{document}